\pdfoutput=1
\documentclass[conference]{IEEEtran}
\IEEEoverridecommandlockouts
\usepackage{cite}
\usepackage{amsmath,amssymb,amsfonts}
\usepackage{algorithmic}
\usepackage{graphicx}
\usepackage{textcomp}
\usepackage{xcolor}
\usepackage{tabularx}
\usepackage{authblk}
\usepackage{subfig}
\usepackage{flushend}

\def\BibTeX{{\rm B\kern-.05em{\sc i\kern-.025em b}\kern-.08em
    T\kern-.1667em\lower.7ex\hbox{E}\kern-.125emX}}

\newlength\mylength
\newlength\mylengthh
\newlength\mylengthhh
\begin{document}

\title{On the Necessity and Design of Coordination Mechanism for Cognitive Autonomous Networks
}



\author[1,2]{Anubhab Banerjee}
\author[1]{Stephen S. Mwanje}
\author[2]{Georg Carle \vspace{-0.28cm}} 
\affil[1]{Nokia Bell Labs, Munich, Germany}
\affil[2]{Dept. of Informatics, Technical University of Munich, Germany}
\affil[ ]{Email:\textit {anubhab.banerjee@tum.de} \vspace{-0.3cm}}

\maketitle

\begin{abstract}
Cognitive Autonomous Networks (CAN) \cite{mwanje2018towards} are promoted to advance Self Organizing Network (SON), replacing rule-based SON Functions (SFs) with Cognitive Functions (CFs), which learn optimal behavior by interacting with the network.
As in SON, CFs do encounter conflicts due to overlap in parameters or objectives. 
However, owing to the non-deterministic behavior of CFs, these conflicts cannot be resolved using rule-based methods and new solutions are required.
This paper investigates the CF deployments with and without a coordination mechanism, and proves both heuristically and mathematically that a coordination mechanism is required. 
Using a two-CF Multi-Agent-System model with the possible types of conflicts, we show that the challenge is a typical bargaining problem, for which the optimal response is the Nash bargaining Solution (NBS).
We use NBS to propose a coordination mechanism design that is capable of resolving the conflicts and show via simulations how implementation of the proposed solution is feasible in real life scenario.

\end{abstract}


\section{Introduction}
\label{intro}

Continuous increase of mobile network users and their online activities have motivated the deployment of several Radio Access Technologies (RATs) to improve spectral efficiency and Quality of Service (QoS).
To address the increasing operational complexity coming from introduction of new RATs, network automation exemplified by Self-Organizing Networks (SON) \cite{hamalainen2012lte} is applied. 
SON proposed to deploy several closed-loop control-based functions, called SON Functions (SFs), each of which deals with a specific problem like Mobility Robustness Optimization (MRO), Mobility Load Balancing (MLB) etc.
Based on the network states (e.g., changes in Key Performance Indicators (KPIs)), an SF determines individual network configuration parameters following some predefined rules. 
Since there are multiple SFs present in a SON, conflicts of interest may arise among them due to overlap of parameters or objectives. 
At the top of these SFs, rule based SON coordination ensures that the SFs do not conflict with one another during operations.


SON has two primary disadvantages -  i) rule based SFs have limited capability in adapting themselves in a changing environment, ii) large number of rules makes maintenance and upgradation of the system difficult.
Cognitive Autonomous Networks (CAN) \cite{mwanje2018towards} overcome the problems of SON and provides a more flexible system by replacing SFs with CFs.
Unlike SF, CF does not exhibit rule-based behavior, rather it learns from the environment and acts based on its learning.
As CFs show a non-deterministic behavior, conflicts among CFs cannot be resolved using existing rule-based methods and some new approach is necessary.
There already exist prior research works which propose different ways on how the CFs can coordinate and work in a decentralized manner \cite{mwanje2015concurrent, mwanje2018synchronized}, but, these works show neither the necessity of collaboration among the CFs nor the optimality of the solution provided.

In this paper our contributions are three-fold - 1. we mathematically prove, using Prisoner's Dilemma from Game Theory \cite{dawes1980social}, that coordination mechanism among the CFs is needed for better performance of the system, 2. we design an easily implementable but effective coordination mechanism capable of determining the optimal configuration for a certain state of the network, and,
3. we also model a heterogeneous Multi Agent System (MAS) in Python that exhibits all types of conflicts \cite{hamalainen2012lte} among the CFs and implement our proposed solution in that model to show the its feasibility in real life.
%

\section{Conflict model in CAN}
\label{conflict_model}

Network automation functions typically exhibit three types of conflicts - 
\begin{itemize}
	\vspace{-0.15cm}\item \textbf{Category A.} Configuration conflict which occurs on either, (A1) Input, or, (A2) Output parameter(s).
	\item \textbf{Category B.} Measurement conflict where action of one function influences measurement of output of another.
	\item \textbf{Category C.} Characteristic conflict which are of two types - (C1)Direct characteristic conflict and (C2)Logical dependency conflict. \vspace{-0.1cm}
\end{itemize}

\begin{figure}[!t]
	\centering
	\includegraphics[scale=0.47, bb=6.858 0.738 347.279474 304.505991]{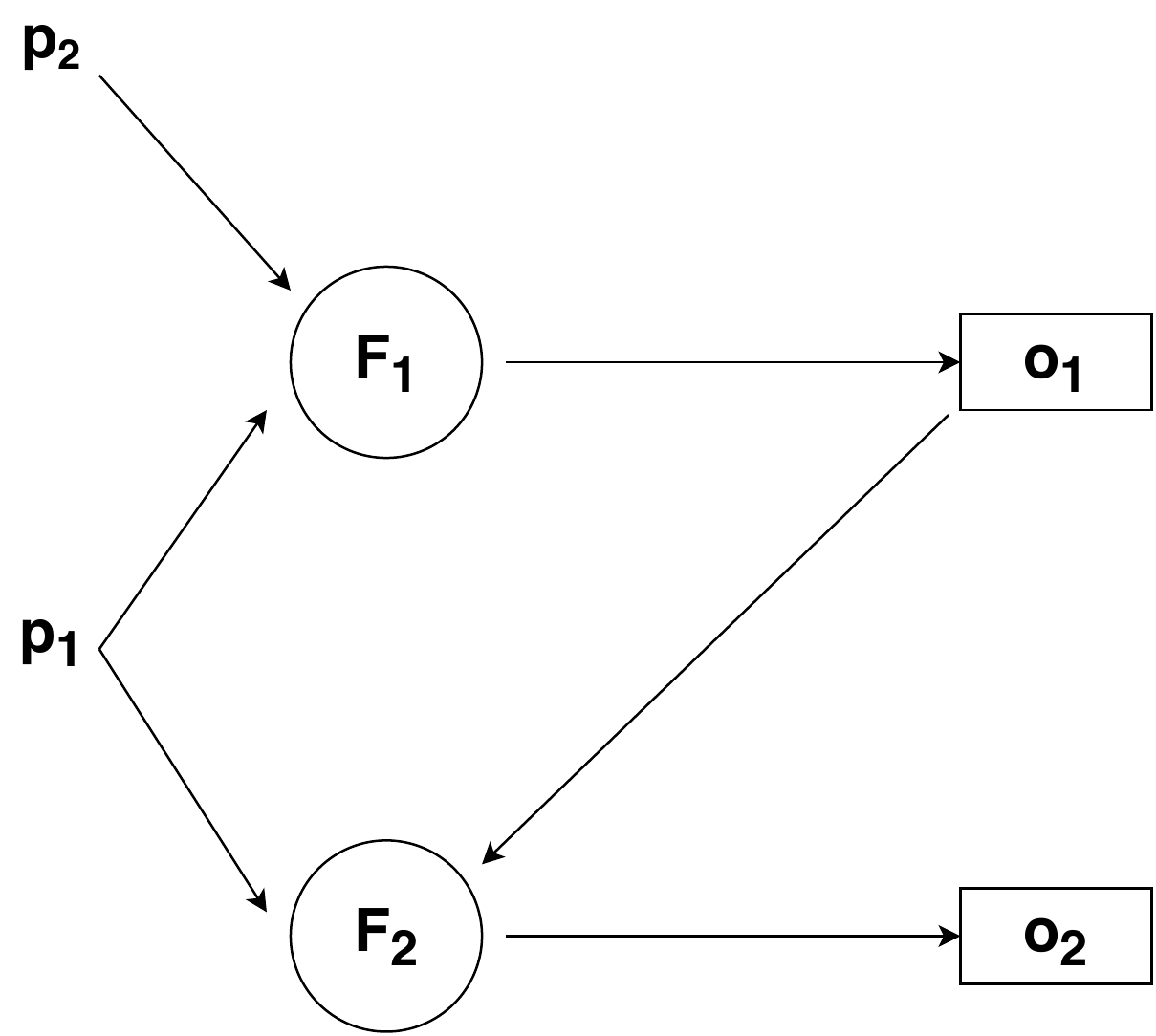}
	\caption{System model}
	\label{img:sys_model}
\end{figure}

Lets us assume that there are two CFs, $F_1$ and $F_2$ working at a Base Station (BS) with objectives $o_1$ and $o_2$ respectively and without communicating with each other.
These are the two agents of the MAS.
There are two inputs to $F_1$ - $p_1$ and $p_2$, and $F_1$ tries to optimize its output $o_1$.
There are two inputs to $F_2$ - $p_1$ and $o_1$, and $F_2$ tries to optimize its output $o_2$.
The system model is depicted in Fig.~\ref{img:sys_model} and the functions with their related information are listed in Table ~\ref{table:functions}.

Now, we can see that all three types of conflicts, are present in this model -
\begin{itemize}
	\item As both $F_1$ and $F_2$ share the same input parameter ($p_1$), so if they have different levels of interest, it is an input parameter conflict (A1).
	\item As actions of $F_1$ affects the measurement of $o_2$, output of $F_2$
	, it is a measurement conflict (B1).
	\item Changing $p_2$ affects $o_1$ which in turn changes $o_2$ hence it is a logical dependency conflict (C2).
\end{itemize}
In the next sections we use this model to prove necessity of coordination mechanism among the CFs and to implement our proposed optimal solution.
\begin{table}[tb]
	\centering
	\begin{tabular}{p{\mylengthh}|p{\mylengthh}|p{\mylengthh}|p{\mylengthh}}
		\hline
		\textbf{Function} & \textbf{Inputs} & \textbf{Outputs} & \textbf{Objective}
		\\ \hline 
		$F_1$ & $p_1$, $p_2$ & $o_1$ & Optimize $o_1$ \\ \hline
		$F_2$ & $p_1$, $o_1$ & $o_2$ & Optimize $o_2$ \\
		\hline 
	\end{tabular}
	\caption{CF Descriptions \vspace{-0.4cm}}
	\label{table:functions}
\end{table}

\section{Necessity of Coordination Mechanism in CAN}
\label{nes_con_can}

Let us consider the model described in section~\ref{conflict_model} and formulate it as a normal-form game \cite{banerjee2019game} where $F_1$ and $F_2$ are the players of the game.
When there is a conflict of interest (A1, B1 or C2) between the players, each player can take one of these following strategies - either, the player continues to work on the interest ($T$), or, it gives up the interest ($G$). 
The payoffs are defined as follows: \vspace{-0.11cm}
\begin{itemize}
	\item Both of them choose $G$: no one changes the interest and it remains constant. Both the CFs get equal payoff out of it denoted by $r_1$.
	\item Both of them choose $T$: 
	the payoff both of them get is $r_2$. The benefit of fighting for the interest is worse than keeping it constant because it may change to worse outcomes for any of the parties and so, $r_2 < r_1$.
	\item One of them selects $T$ and the other selects $G$: one who selects $T$ gets a payoff $r_3$ and the other one gets a payoff $r_4$ with $r_3 > r_4$. \vspace{-0.11cm}
\end{itemize}
\textit{Relation among $r_1$, $r_2$, $r_3$ and $r_4$}:
It is obvious that payoff is higher when the CF can adjust the interest than when it remains constant, i.e., $r_3 > r_1$. 
The payoff for one is also higher when either the interest is controlled by itself or the interest remains constant than when it is changed according to the other's will, i.e., $r_2 > r_4$ and $r_1 > r_4$. 
The payoff for a CF is higher again when only the CF changes it than when both of them change it, i.e., $r_3 > r_1$. 
The above observations can be summarized as: \vspace{-0.28cm} 
\begin{equation}
r_3 > r_1 > r_2 > r_4   
\label{eq:payoff_relations}
\vspace{-0.099cm}
\end{equation}

As mentioned in \cite{dawes1980social}, two criteria for a problem to qualify as \textit{Prisoner's Dilemma} 
are: \vspace{-0.11cm}
\begin{itemize}
	\item regardless of what the other players do, each player receives a higher payoff for defecting behavior than for cooperating behavior.
	\item all agents get lower payoff if all defect than cooperate. \vspace{-0.11cm}
\end{itemize}
Now, a conflict between CFs can be formulated exactly as a Prisoner's Dilemma where the defecting behaviors is choosing $T$ and cooperative behavior is choosing $G$, because, from Eq.~\ref{eq:payoff_relations}, we observe that: \vspace{-0.11cm}
\begin{itemize}
	\item Regardless of what the other CF does, each CF receives a higher payoff for selecting $T$ than selecting $G$.
	\item All CFs receive a lower payoff if all choose $T$ than if all choose $G$. \vspace{-0.11cm}
\end{itemize}
Following the solution of Prisoner's Dilemma, where best action for each prisoner is to choose defecting behavior, the best action for each CF is: \textbf{ selecting $T$}. 

On the contrary, if the CFs work with the existence of a coordinating mechanism, they find that the best possible action for each CF is - \textbf{choosing $G$}, because, when both of them select $G$, both of them get higher payoff ($r_1$) than the payoff they get ($r_2$) when they select $T$.
This proves the necessity of a coordination mechanism in the network.

\section{Proposed Solution and Implementation}
\label{proposed_solution}

\begin{figure*}[htp]
	\vspace{-0.6cm}	
	\centering	
	\subfloat[Variation of $o_1$ against $p_1$ and $p_2$]{%
		\includegraphics[clip,width=0.9\columnwidth]{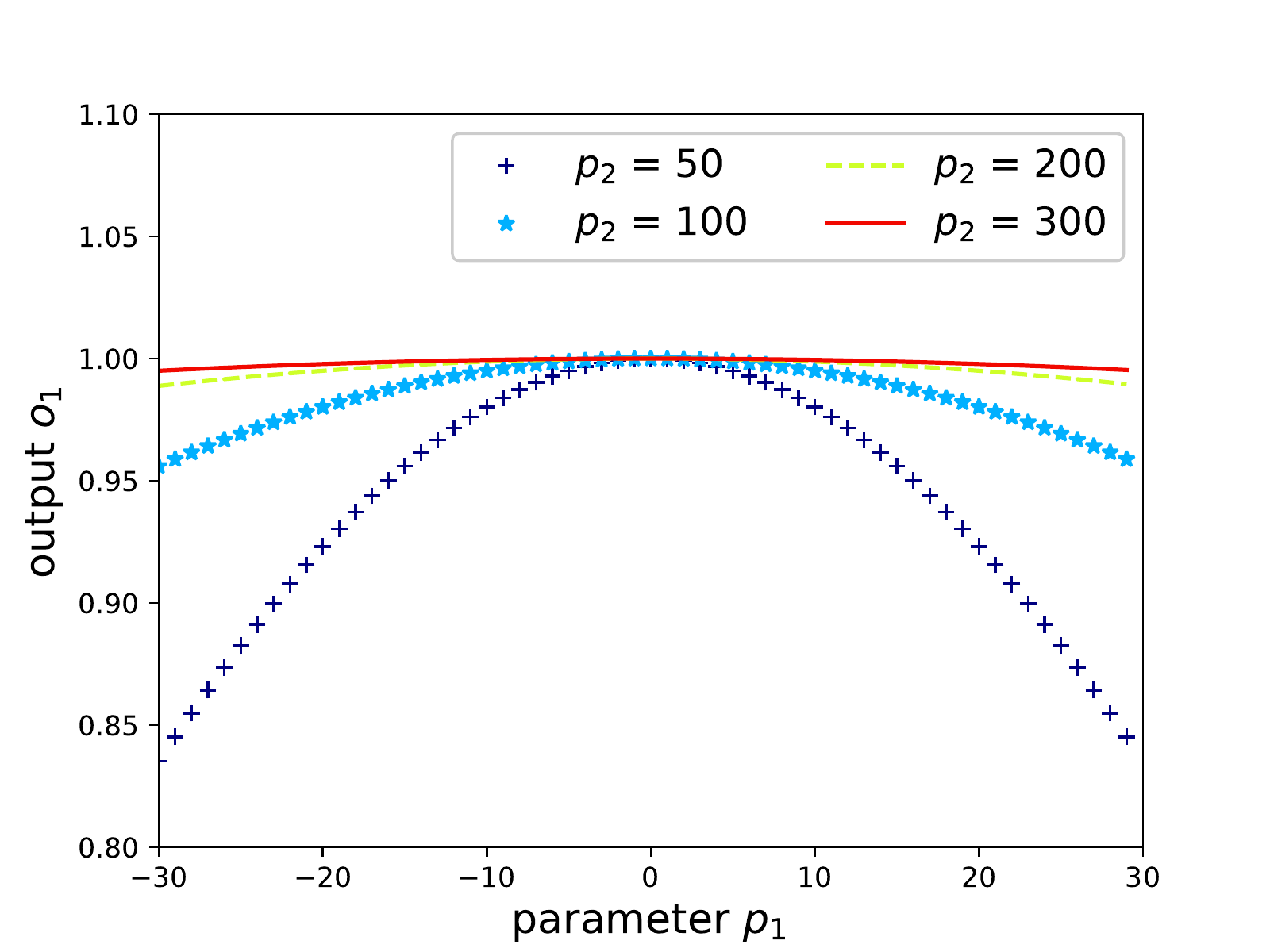}%
		\label{img:o1_p1p2}
	}	
	\subfloat[Variation of $o_2$ against $p_1$ and $p_2$]{%
		\includegraphics[clip,width=0.9\columnwidth]{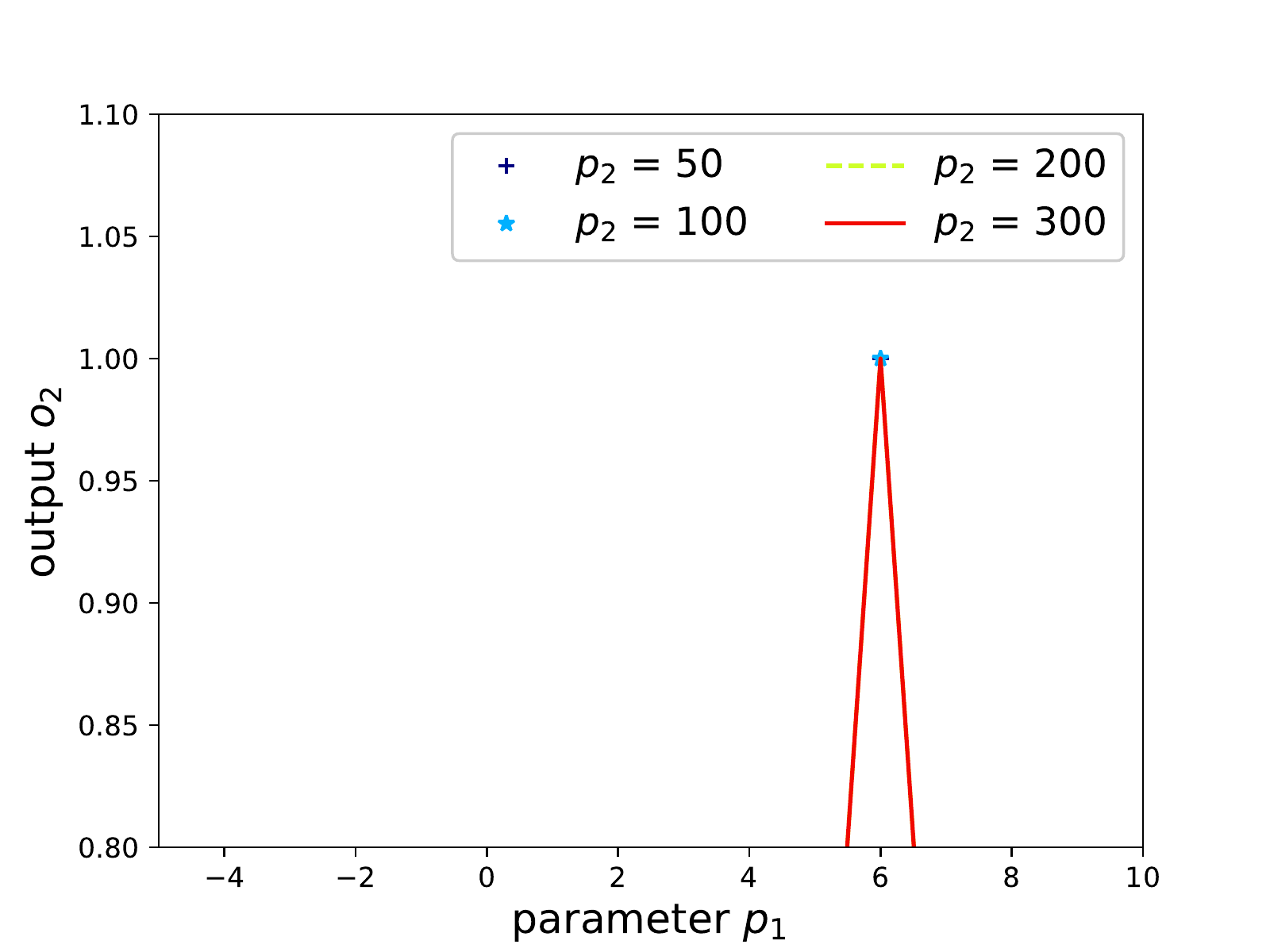}%
		\label{img:o2_p1p2}
	}	
	\caption{Output variations against input parameters \vspace{-0.28cm}}
	\label{img:p1p2}	
\end{figure*}
\begin{figure*}[htp] 
	\vspace{-0.6cm}	
	\centering
	\subfloat[Variation of $o_1 \cdot o_2$ against $p_1$]{%
		\includegraphics[clip,width=0.9\columnwidth]{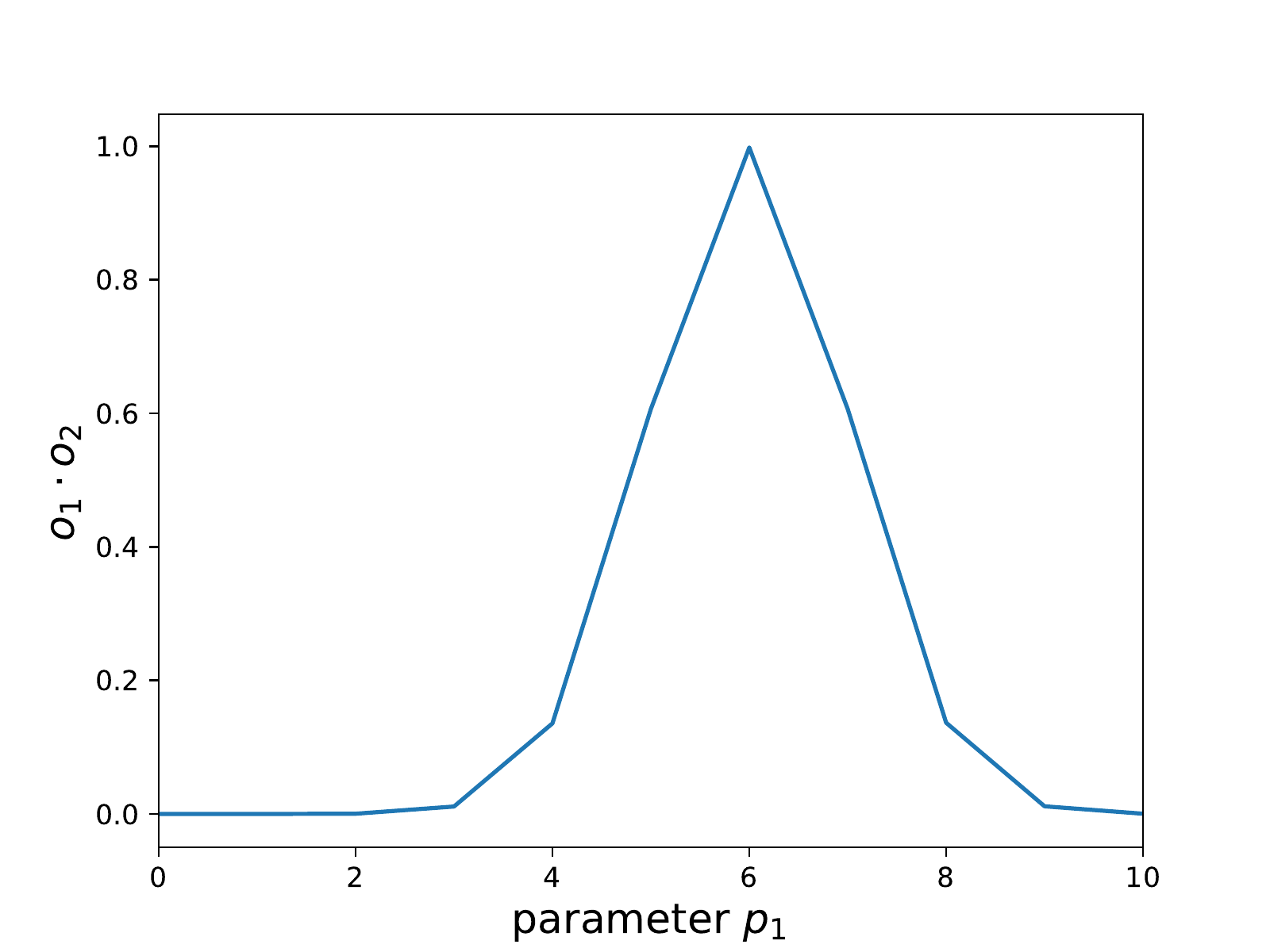}%
		\label{img:o1o2_p1}
	}	
	\subfloat[Variation of $o_1 \cdot o_2$ against $p_2$]{%
		\includegraphics[clip,width=0.9\columnwidth]{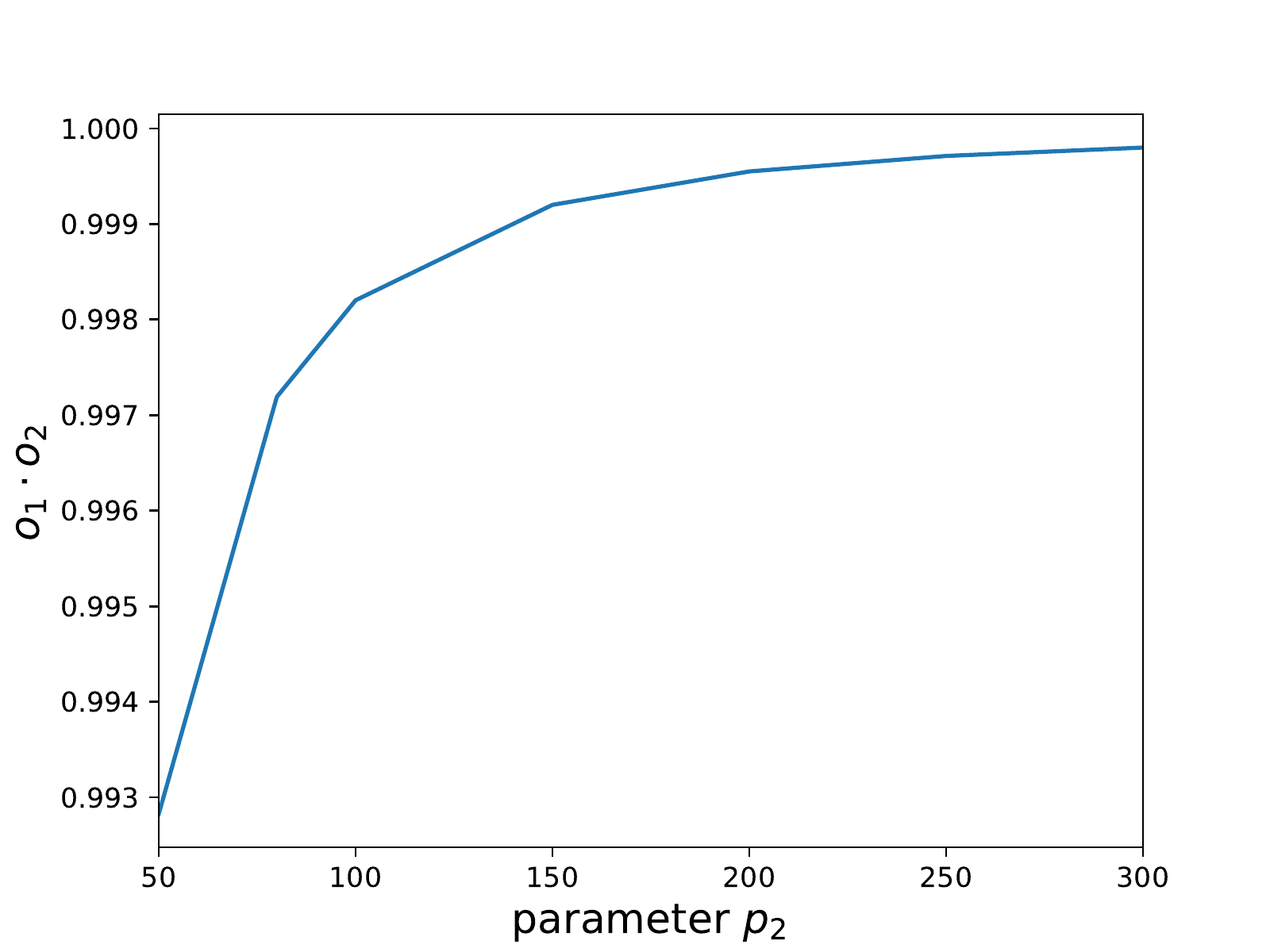}%
		\label{img:o1o2_p2}
	}	
	\caption{Variation of $o_1 \cdot o_2$ against input parameters \vspace{-0.28cm}}	
	\label{img:o1o2}
\end{figure*}
Let us again consider the model described in section~\ref{conflict_model} and assume that both $F_1$ and $F_2$ try to maximize their objectives and both $o_1$ and $o_2$ are Gaussian distribution functions given by the following equations: 
\begin{equation}
o_1 = e^{- \frac{p_1^2}{2 p_2^{2}}}
\label{eq:o1}
\end{equation}
\begin{equation}
o_2 = e^{- \frac{(p_1 - 6)^2}{\frac{2}{o_1^{2}}}} 
\label{eq:o2}
\end{equation}
We also assume that default value (value at the initial stage when BS first becomes operational) of $p_1$ is 4 and $p_2$ is 100.
These functions ($o_1$, $o_2$) and values ($p_1$, $p_2$) are assumed in such a way so that all types of conflicts are present between $F_1$ and $F_2$.

\subsection{Proposed solution}
\label{pr_sol}

To resolve the conflicts between $F_1$ and $F_2$ over the values of $p_1$ and $p_2$, we propose a bargaining between them and obtain the solution using Nash Bargaining Solution (NBS) \cite{binmore1986nash}.
NBS can be applied when - i) number of players is two or more, ii) there is a conflict of interest, and iii) there exists a solution if  if negotiation breaks down, and the solution is obtained by finding the value for which product of outcome of the players is maximum.
As the number of persons is two and there is a conflict of interest on agreement, Nash Bargaining Solution (NBS) can be applied to obtain the solution in this scenario \cite{banerjee2019game} and this solution is optimal \cite{binmore1986nash}.
The optimal value of $p_1$ is determined following these steps sequentially: \vspace{-0.11cm}
\begin{itemize}
	\item Both $F_1$ and $F_2$ generate a set of possible values for $p_1$ based on their learning history. 
	\item For each value of $p_1$ in this set, the product of $o_1$ and $o_2$ is calculated while $p_2$ is kept constant.
	\item When the product is maximum, the corresponding value of $p_1$ is the optimal value. \vspace{-0.11cm}
\end{itemize}
After the optimal value of $p_1$ is determined, a set of possible values for $p_2$ is also generated by $F_1$ based on its learning history.
For each value of $p_2$ in this set, the product of $o_1$ and $o_2$ is calculated while $p_1$ is kept constant at its optimal value.
When the product is maximum, the corresponding value of $p_2$ is the optimal value.
In this way the optimal configuration for the system can be determined.

In this solution we assume that $F_1$ and $F_2$ can generate sets of values for $p_1$ and $p_2$ based on their previous learning history.
As an alternate, starting from the default, values of these parameters can be changed slowly and gradient descent approach can be used to reach the optimal.
\begin{table*}[htb]
	\centering
	\begin{tabular}{p{\mylength}|p{\mylength}|p{\mylength}|p{\mylength}|p{\mylength}|p{\mylength}|p{\mylength}|p{\mylength}|p{\mylength}|p{\mylength}|p{\mylength}|p{\mylength}|p{\mylength}|p{\mylength}|p{\mylength}|p{\mylength}}
		\hline \vspace{0.001cm}  
		\textbf{F1} & \checkmark & & & & \checkmark & \checkmark & \checkmark & & & & \checkmark & \checkmark & \checkmark & & \checkmark \\ \hline \vspace{0.001cm}
		\textbf{F2} & & \checkmark & & & \checkmark & & & \checkmark & \checkmark & & \checkmark & \checkmark & & \checkmark & \checkmark  \\ \hline \vspace{0.001cm}
		\textbf{F3} & & & \checkmark & & & \checkmark & & \checkmark & & \checkmark & \checkmark & & \checkmark & \checkmark & \checkmark \\ \hline \vspace{0.001cm}
		\textbf{F4} & & & & \checkmark & & & \checkmark & & \checkmark & \checkmark & & \checkmark & \checkmark & \checkmark & \checkmark  \\ [0.5ex] 
		\hline \hline
		& \vspace{0.005mm}\cite{sarika2015agenttab} 
		& \vspace{0.005mm}\cite{khayyat2016an} 
		& \vspace{0.005mm}\cite{rahimzadeh2015high} 
		& \vspace{0.005cm}\cite{banerjee2019sharing}
		& \vspace{0.005mm}\cite{mano2004self} 
		& \vspace{0.005mm}\cite{gatti2013large} 
		& \vspace{0.005mm}\cite{manvi2008multicast} 
		& \vspace{0.005mm}\cite{morstyn2015cooperative} 
		& \vspace{0.005mm}\cite{bianchi2013heuristically} 
		& \vspace{0.005cm}\cite{resmerita2003conflict}
		& \vspace{0.005mm}\cite{semsar2009game} 
		& \vspace{0.005mm}\cite{gupta2017cooperative} 
		& \vspace{0.005cm}\cite{liu2015novel}
		& \vspace{0.005mm}\cite{liu2008conflict} 
		& \vspace{0.005mm}\textbf{X} \\ [1ex] 
		\hline 
	\end{tabular}
	\vspace{0.21cm}
	\caption{Existing works on MAS features \vspace{-0.4cm}}
	\label{table:related-works}
\end{table*}

\subsection{Implementation and Observation}
\label{eval}

We build a framework in Python to implement the proposed solution in the MAS discussed in section~\ref{conflict_model}.

In Fig.~\ref{img:p1p2} variations of $o_1$ and $o_2$ against $p_1$ and $p_2$ have been plotted.
From Fig.~\ref{img:o1_p1p2} we see that for a fixed $p_2$, maximum value of $o_1$ is obtained when $p_1$ lies in between -10 and 10 and maximum value of $o_2$ is obtained when $p_1$ lies in between 0 and 10.
Thus, in our simulation, we vary $p_1$ in between 0 and 10 in steps of 1 (step size can be made smaller for better accuracy) and plot the variation of $o_1 \cdot o_2$ against $p_1$ in Fig.~\ref{img:o1o2_p1}.
From Fig.~\ref{img:o1o2_p1} we see that $o_1 \cdot o_2$ is maximum when $p_1$ is 6, and so, according to NBS, this is the optimal value of $p_1$ for both $F_1$ and $F_2$. 

Once the optimal value of $p_1$ is determined, it is kept constant at 6 and $p_2$ is varied to determine its optimal value.
For a fixed $p_1$, when the value of $p_2$ increases, we see from Fig.~\ref{img:o1_p1p2} that the value of $o_1$ increases whereas from Fig.~\ref{img:o2_p1p2} we see that the value of $o_2$ remains almost constant. 
For this reason, when we plot $o_1 \cdot o_2$ against $p_2$ in Fig.~\ref{img:o1o2_p2} we see that $o_1 \cdot o_2$ increases with increase in value of $p_2$.
As a specific range for $p_2$ cannot be determined, we vary $p_2$ between 50 and 300 in our simulation and according to NBS, optimal value of $p_2$ in this scenario is 300.
However, in real life there is always a maximum and minimum value for a parameter between which it can be varied.

\section{Related Works and Motivation}
\label{related_works_and_motivation}

As the CAN with CFs has been formulated as a MAS in this paper, in this section we discuss relevant existing research works on MAS (\cite{liu2008conflict, gatti2013large} ) and removal of conflicts (reaching a consensus) in a MAS (\cite{resmerita2003conflict, genesereth1988cooperation} ).

In our MAS model, described in section~\ref{conflict_model}, each agent, i.e. CF, has the following features: 
\begin{itemize}
	\item[F1.] each agent can learn and decide what is the best action for it by itself in a dynamic environment.
	\item[F2.] no agent can communicate with each other and no one has a complete knowledge of the system. 
	\item[F3.] some or all of these agents share the same resources and there exist conflicts of interests among them.
	\item[F4.] each agent tries to optimize its own target or goal simultaneously, and the concept of a common or team goal does not exist.
\end{itemize}

Based on agent characteristics, we divide existing research works on MAS into several categories so that a combination of these features are covered in each category.
These categories are listed in Table~\ref{table:related-works}.
From Table~\ref{table:related-works} we see that there are a number of prior research articles which encompass one or some combinations of those four features described above, but there does not exist any paper which covers all the four features (as shown in Table~\ref{table:related-works}).
Ours is the first one which considers a MAS with all of these four properties.


\section{Conclusion and Future Direction}
\label{conclusion}

In this paper we discuss conflict resolution among cognitive functions and prove, using Prisoner's Dilemma, that a coordinating mechanism is needed for better performance of the system.
We provide a solution to resolve the possible types of conflicts among the CFs.
We build a MAS in Python and show that our proposed solution can be implemented there to resolve the conflicts and obtain the optimal settings for the operational purposes.
Currently we are building prototypes of these cognitive functions and as a next step we want to test them in a simulation environment to do a comparative study among possible coordination mechanisms.

\bibliographystyle{unsrt}
\bibliography{references}

\end{document}